# Using an Isomorphic Problem Pair to Learn Introductory Physics: Transferring from a Two-step Problem to a Three-step Problem

Shih-Yin Lin and Chandralekha Singh
Department of Physics and Astronomy, University of Pittsburgh,
Pittsburgh, PA 15260, USA

**Abstract**

In this study, we examine introductory physics students' ability to perform analogical reasoning between two isomorphic problems which employ the same underlying physics principles but have different surface features. Three hundred and eighty two students from a calculus-based and an algebra-based introductory physics course were administered a quiz in the recitation in which they had to learn from a solved problem provided and take advantage of what they learned from it to solve another isomorphic problem (which we call the quiz problem). The solved problem provided has two sub-problems while the quiz problem has three sub-problems, which is known to be challenging for introductory students from previous research. In addition to the solved problem, students also received extra scaffolding supports that were intended to help them discern and exploit the underlying similarities of the isomorphic solved and quiz problems. The data analysis suggests that students had great difficulty in transferring what they learned from a 2-step problem to a 3-step problem. Although most students were able to learn from the solved problem to some extent with the scaffolding provided and invoke the relevant principles in the quiz problem, they were not necessarily able to apply the principles correctly. We also conducted think-aloud interviews with six introductory students in order to understand in-depth the difficulties they had and explore strategies to provide better scaffolding. The interviews suggest that students often superficially mapped the principles employed in the solved problem to the quiz problem without necessarily understanding the governing conditions underlying each principle and examining the applicability of the principle in the new situation in an in-depth manner. Findings suggest that more scaffolding is needed to help students in transferring from a two-step problem to a three step problem and applying the physics principles appropriately. We outline a few possible strategies for future investigation.

## I. INTRODUCTION

Identifying the relevant physics principles involved is one important component of problem solving in physics. Physics is a subject in which diverse physical phenomena can be explained by just a few basic physics principles. To learn physics effectively, it is essential to unpack the meaning of the abstract principles, and understand the applicability of the physics principles in diverse situations [1-9]. One major goal of many physics courses, therefore, is to help students learn to discern the deep similarities between the problems that share the same underlying physics principles but have different surface features, so that students can transfer what they learn from one context to another. However, it is well known that two physics problems that look very similar to a physics expert because both involve the same physics principle don't necessary look similar to the beginning students [1,10]. On the other hand, problems that the beginning students consider as similar may actually involve very different physics concepts in the solution steps. For example, a study on the categorization of introductory mechanics problems [1] based upon similarity of solutions indicates that while experts are likely to place two problems in different categories because one of the problems involves one physics principle (e.g., the principle of conservation of energy) but the other problem involves a different principle (such as Newton's 2nd Law), novices may group two problems together because both of them involve an inclined plane. The findings suggest that experts usually group problems based upon the physics principles while novices are more likely to be distracted and group the problems based on the surface features (such as the inclined plane or pulley). The different ways experts and novices categorize problems also reflect the different ways knowledge is organized in their minds [1,2,11-19]. Research suggests that experts in physics have a highly hierarchical knowledge structure, where the most fundamental physics principles are placed at the top, followed by layers of subsidiary knowledge and details [1,10,20]. This well-organized knowledge structure facilitates their problem solving process and helps them approach the problems in a systematic way [1,2,11-19,21,22]. It also guides the experts to see the problems beyond the surface features, and makes the transfer of knowledge between different contexts easier. As novice students' knowledge structure is usually less organized, it will be beneficial if the instruction provided can help them construct a robust, well-organized knowledge hierarchy (e.g., by learning to extract the deep connections between problems that share the same underlying physics concepts,) and to understand the broad applicability of the overarching physics principle in various contexts.

Helping students apply what he/she has learned in one situation to a different situation is an important goal of education. Therefore, a lot of research efforts have been devoted to investigating transfer of learning. In these investigations, issues about transfer of learning have been widely discussed from different perspectives [19,23-31]. For example, the degree to which students can apply knowledge flexibly [23,26,29,32-37], learning features that affect transfer [37-42], and the possible framework to characterize transfer [26-30,33-35,41,43] are discussed in various research contexts. It is pointed out that the amount of knowledge a person has, the knowledge structure that the person constructs, and the context in which the knowledge is learned can all affect the person's ability to apply knowledge flexibly [29]. In order to assist students in learning and help them transfer their learning to different contexts, various scaffolding mechanisms can be used. For example, students can be taught to perform analogical reasoning [1,2,37-39,41,44-46] between problems that involve the same underlying physics principles. Studies have shown that using analogy can help improve students' learning and reasoning in many domains [47-52]. A good analogy can help people understand an unfamiliar situation more easily by creating a connection between the new and existing information [53]. Such connection can make the mental processing of new information more efficient by modifying the existing knowledge schemata. It can also make the new information more concrete and easier to comprehend. Analogy has long been an effective strategy adopted by many instructors in the classrooms. It is also a common practice for students to solve new problems by first looking for similar problems that they already know how to solve, and applying similar reasoning strategies from one problem to another. To help students recognize the applicability of a physics principle in various contexts by performing analogical problem solving, students can be explicitly guided to point out the similarities between two problems that involve different surface features but the same underlying physics, and take advantage of what they learn from one problem to solve the other. In doing so, students may develop an important skill shared by experts: the ability to transfer from one context to another, based upon shared deep similarities.

In this study, we examine students' abilities to learn from worked out examples and to perform analogical problem solving between two isomorphic problems. In particular, we investigate if students could discern the similarities between a solved problem and a quiz problem, take advantage of the similarities and transfer what they learn from the solved problem to solve the quiz problem that is isomorphic. According to Hayes and Simon's definition [54], problems are isomorphic if they can be mapped to each other in a one-to-one relation in terms of their solutions and the moves in the problem solving trajectories. For example, the "tower of Hanoi problem" is isomorphic to the "cannibal and the missionary problem" since they have the same structure if reduced to the abstract mathematical form [54]. In our study, we call problems isomorphic if they can be solved using the same physics principles. Research has shown that two problems which are isomorphic are not necessarily perceived as being at the same level of difficulty, especially by a beginning learner [55,56]. Depending on a person's expertise in the field, different contexts and representations may trigger the recall of a relevant principle more in one problem than another. Changing the context of the problem, making one problem in the isomorphic pair conceptual and the other quantitative, or introducing distracting features into one of the problems can to different extent raise the difficulty in discerning the similarity and make the transfer of learning between the two problems more challenging [57].

In a prior study [46], we have investigated students' abilities to transfer their learning from a 2-step solved problem to a 2-step isomorphic quiz problem. In particular, students were explicitly asked in a recitation quiz to browse through and learn from a solved problem (to which a detailed solution was provided) and then use the analogy to solve an isomorphic quiz problem. In that prior study [46], the solved problem provided was about a girl riding a roller coaster on a smooth track. The roller coaster car was initially at rest at a certain height. The problem asked for the apparent weight of the girl as the roller coaster car went over the top of a circular hump given the girl's weight, the radius of the circle, and the heights of different points. The quiz problem, on the other hand, was about a boy on a tire swing created with a rope tied to a tree. Students were asked to find the maximum tension in the rope during the ride given the boy's mass, the length of the rope, and the initial height assuming the boy starts from rest at the initial height. Although these two problems may look very different to a novice student, the solutions to the solved and quiz problems can be matched to each other in a one-to-one manner. Both of them can be solved by decomposing them into two sub-problems and applying the principles of conservation of mechanical energy and Newton's $2^{nd}$ Law with centripetal acceleration in each subproblem, respectively. Different types of scaffolding (instructional support) were provided to students in different intervention groups in order to assist students in transferring the knowledge they learned from the solved problem to the quiz problem. Although the quiz problem was challenging, the prior study [46] found that with the proper scaffolding support provided, students were able to reason through the analogy between those two problems and performed significantly better on the quiz problem (the tire swing problem) than students who were not provided with the isomorphic solved problem (the roller coaster problem) to learn from.

In this study, students' ability to perform analogical problem solving and transfer their learning from one situation to another between another pair of isomorphic problems is explored. Unlike the previous study [46] in which both the solved and quiz problems are 2-step problem, the quiz problem in this study can be solved by decomposing it into three sub-problems while the solved problem requires decomposing it into two sub-problems. Since prior research [58] indicated that many students struggle with this 3-step quiz problem, the goal in our current study was to examine whether students can benefit from the solved problem and other scaffolding provided, in a case in which the quiz problem involves one more step than the solved problem and can no longer be directly mapped to the solved problem without careful thinking.

## II. METHODOLOGY

### A. The analogical problem solving activity and the problems

In this study, students from a calculus-based and an algebra-based introductory physics course were given two isomorphic problems in the recitation quiz. The solution to one of the problems (which we call the "solved problem") was provided. Students were explicitly asked to learn from the solution to the solved problem, point out the similarities between the two problems, explain whether and how they can exploit the solved problem to solve the other problem (which we call the "quiz problem"). Then they were asked to solve the quiz problem. The solution provided was presented in a detailed and systematic way. It started with a description of the problem with the knowns, unknowns, and target quantity listed, followed by a plan for solving the problem in which the reasons for why each principle was applicable were explicated. After the plan was executed in the mathematical representation, the last part of the solution provided a check for the answer by examining the limiting cases. The solved problem (with its full solution that was provided to the students) and the quiz problem can be found in Appendix A and B, respectively.

The solved problem was about a boy who took a running start, jumped onto a stationary snowboard and then went up a hill with the snowboard. The problem asked for the minimum speed at which the boy should run (right before jumping onto the snowboard) in order to go up to a certain height assuming the frictional force can be neglected. The quiz problem, on the other hand, was about two putty spheres hanging on massless strings of equal length. Sphere A was raised to a height $h_o$ while keeping the string straight. After it was released, it collided with the other sphere B, which has the same mass; the two spheres then stuck and swung together to a maximum height $h_f$. Students were asked to find $h_f$ in terms of $h_o$. Both the solved and quiz problems involve an inelastic collision and process(es) in which something goes up or down while there's no work done by non-conservative forces. Both problems can be solved using the principles of conservation of momentum (CM) and conservation of mechanical energy (CME). However, the snowboard problem can be solved by decomposing it into two steps (first the inelastic collision process, which involves the CM principle, followed by the process of the person and snowboard together going up the hill, which requires the CME principle) while the putty problem involves a three-step solution (with the CME, CM, and CME principles applicable to the processes of putty A going down, inelastic collision, and putties A and B together going up to a maximum height, respectively.) Unlike the previous study [46], in which both the solved and quiz problems are two-step problems and the solutions can be mapped directly to each other, in this study, only the last two steps of the quiz problem and not the whole problem can be mapped directly to the solution of the solved problem. We note that even though the two problems may look very similar to a physics expert and both are relatively easy for them, our previous research indicates that the three-step putty problem is typically very challenging for the introductory students [59]. The investigation in this study was designed to investigate the extent to which providing different types of scaffolding support to students to think about the similarities between the solved problem and the quiz problem may facilitate transfer of what they learned from the two-step solved problem to solve the three-step quiz problem.

## B. Participants and the different interventions

One hundred and eighty students from a calculus-based introductory physics course and 202 students from an algebra-based introductory physics course were involved in this study. In each of the courses, students were randomly divided into one comparison group and three intervention groups based on different recitation classes. There was no significant difference between any of the group in each course in terms of students' force concept inventory (FCI) score conducted at the beginning of the semester.

Students in the comparison group were given only the quiz problem in the recitation quiz. Similar to a traditional quiz, students in this comparison group were asked to solve the quiz problem on their own in 15 minutes; no scaffolding support was provided. The performance of this group of students could help us understand what students were able to do without being explicitly provided a solved isomorphic problem to learn from.

Students in the three intervention groups, on the other hand, were given an opportunity to learn from the solved isomorphic problem during the quiz. As research on learning from worked-out examples [60-64] suggests, larger learning gain can be achieved if students are actively engaged in the process of sense making while learning from examples. In order to help students process through the analogy between the two problems deeply and to contemplate issues which they often have difficulty with, different kinds of scaffolding supports were provided in addition to the solved problem to the students in different intervention groups. A summary of the different scaffolding supports implemented in each intervention group is presented in Table 1. We will discuss the details and the rationale behind each intervention in the following paragraphs.

In particular, students in the intervention group 1 were asked to take the first 10 minutes in the quiz to learn from the solution to the solved problem (the snowboard problem). They were explicitly told at the beginning of the quiz that after 10 minutes, they had to turn in the solution, and then solve two problems in the quiz: one of them would be exactly the same as the one they just browsed over (the snowboard problem), and the other one would be similar (the putty problem.) In order to help students discern the connection between the two problems, students were also explicitly asked to identify the similarities between the two problems and explain whether they could use the similarities to solve the quiz problem before actually solving it. The fact that the solution we provided made explicit the consideration for using the principles but was not directly the solution to the quiz problem was inspired by Schwartz, Bransford and Sears' theory of transfer [65], which states that two components -efficiency and innovation- are both important in the learning and transfer processes. We hypothesized that since students had to solve the same problem they browsed over (i.e., a task toward the efficiency domain) and an isomorphic problem in the quiz (i.e., a task toward the innovation domain), students would try their best to get the most out of the solution in the allocated learning period. In order to apply what they learned from the solution to solve exactly the same problem on their own as well as the isomorphic problem, they had to not only figure out what principles to use, but also understand why and how each principle is applicable in different circumstances. We hypothesized that an advantage could be achieved over the comparison group if students in the intervention group 1 went through a deep reasoning while browsing over the solved problem as we intended. Students' performance on both problems was later analyzed.

**Table 1. Summary of the different interventions used in this study.**

| Group name | Students were asked to…. | Compared to the interventions used in the prior study [46]… |
|---|---|---|
| Comparison | Solve the quiz problem on their own. No solved problem was provided. | The same |
| Intervention 1 | (a) First learn from the solved problem provided<br>(b) Return the solution to the solved problem<br>(c) Solve both the solved and quiz problems | The same |
| Intervention 2 | (a) Solve the quiz problem on their own first<br>(b) Learn from the solution to the solved problem<br>(c) Redo the quiz problem a second time (with the solved problem in their possession) | The same |
| Intervention 3 | Learn from the solution to the solved problem and then solve the quiz problem (with the solved problem in their possession). They were also given extra hints about (a) the fact that similar principles (CM and CME) can be used to solve both problems (b) they might have to use CME twice.<br><br>Note: The exact wording for this intervention can be found in Appendix B. | The same except that the problem-specific hints (a) and (b), which are designed based on common student mistakes on the quiz problem are modified<br><br>Note: In the prior study [46], students in the intervention group 3 were asked to: Learn from the solution to the solved problem and then solve the quiz problem (with the solved problem in their possession). They were also given extra hints about (a) the fact that similar principles (CME and Newton's second law with centripetal acceleration involved) can be used to solve both problems (b) the implications of two common conceptions of centripetal force (one is correct and the other corresponds to a common student mistake). Students were also guided to select the conception they agreed with and discuss why. |

The scaffolding in the intervention group 2 was designed based on a different framework. Students in this group were first asked to solve the quiz problem on their own. After a designated period of time, they turned in their solutions, and were given the isomorphic solved problem to learn from. Then, with the solved problem and its solution in their possession, they were asked to redo the quiz problem a second time after pointing out the similarities between the two problems and explicitly asked to discuss the implication of these similarities in constructing their solution to the quiz problem. We hypothesized that postponing the browsing over the solved isomorphic problem until the students have actually tried to solve the quiz problem on their own could be beneficial to them because in this way, students would have already searched through their knowledge base of physics and attempted to organize the information given in the quiz problem. We hypothesized that having tried the quiz problem on their own may make the browsing over the solved problem for relevant information more structured and productive before students attempted the quiz problem a second time. Even if their initial method of solution was incorrect or couldn't lead them very far, the thinking processes involved may still provide a useful framework for interpreting, incorporating and accommodating the material that they later learned from the solved problem. We hypothesized that if they got stuck in the first trial without scaffolding, this initial struggle and then browsing over the solved isomorphic problem would give them some perspective on why they were stuck and they may become more deliberate and directed in terms of what to look for in the solution. If they failed to recall a certain principle or forgot to take into consideration a certain part in the problem, the similarity between the two problems may trigger the recall of the previously inaccessible knowledge resource. Moreover, if students were not sure whether their solution was correct, the comparison between the two solutions (one provided, one their own) could also serve as a basis for examining the correctness of their answers. Students had the opportunity to display what they learned from the solved isomorphic problem when they solved the quiz problem a second time.

Unlike the students in the intervention groups 1 and 2 who had to figure out the similarities between the two problems themselves, students in the intervention group 3 were given a different type of hint in the quiz. They were given both the quiz problem and the solved problem at the same time. In addition to the instruction which asked them to first learn from the solved problem and then exploit the similarity to solve the quiz problem, students were explicitly told that "Similar to the solved problem, the quiz problem can be solved using conservation of momentum and conservation of mechanical energy." We hypothesized that deliberately pointing out that similar principles should be used in both problems may guide students to focus more on the deep physics principles. Moreover, students in this group were explicitly told that they may have to use the conservation of energy twice because our previous research indicates that it's challenging for students to recognize the three-step nature of the putty problem [58]. The full instruction provided to intervention group 3 can be found in Appendix B.

In order to facilitate discussion of our research findings, a comparison between the interventions used in the prior study [46] and our current study are presented in Table 1. Except for intervention 3, in which some of the additional instructions provided are problem specific (e.g., the instruction of "applying the CME twice" is designed based on the common student difficulties on the quiz problem that were found in the prior research), the other interventions used in our current study were similar to those used in the prior study [46] in which students were asked to transfer between a pair of isomorphic problems. However, even for interventions 1 and 2, the current study investigates transfer from a two part problem to a three part problem unlike the prior study where both the solved and transfer problems were two part problems.

## C. Data Analysis

In order to examine the effects of the scaffolding supports, students' performance in the quiz was scored by two researchers on a rubric that was developed by both researchers. We found that the similarities between the solved and quiz (transfer) problems that the students described in the first part of their quiz solution didn't provide useful information about their ability to actually solve the quiz problem. Common similarities that the students recognized include following observations: both problems involve an inelastic collision, the principle of conservation of mechanical energy can be used etc. However, the students didn't necessarily point out how the quiz (transfer) problem can be broken into different sub-problems and in which sub-problem should each principle be applied. Therefore, in the following discussion, we will only focus on students' *solutions* to the quiz problem.

Students' solutions to the quiz problem were scored using the rubric developed. Summaries of the rubrics for the solved problem and the quiz problem are shown in Table 2 and Table 3, respectively. The rubrics consist of 2 parts based upon the principles required. Different scores were assigned in the solved problem than in the quiz problem because the former involves a 2-step solution and the latter involves 3 steps. The rubrics were designed taking into account the common student difficulties found. An inter-rater reliability of more than 80 percents was achieved when two researchers scored independently a sample of 20 students.

**Table 2.** Summary of the rubric for the solved problem.

| Description | Scores |
|---|---|
| Conservation of Momentum (5 points) | Invoking physics principle: 3 points (Students received a score of either 3 or zero, depending on whether the physics principle was invoked in the solution) |
| | Applying physics principle: 2 points (One point was taken off if students made a mistake such as plugging in the wrong number for the masses) |
| Conservation of Mechanical Energy (5 points) | Invoking physics principle: 3 points (Students received a score of either 3 or zero, depending on whether the physics principle was invoked in the solution) |
| | Applying physics principle: 2 points (One point was taken off if they made a mistake such as plugging in the wrong number for height or mistakenly writing the kinetic energy as ½ $mv$ or $mv^2$, where $m$ stands for the mass, $v$ stands for the speed of the object(s) of interest ) |

Table 3. **Summary of the rubric for the quiz problem.**

| Description | Scores |
|---|---|
| Conservation of Mechanical Energy in the 1st and 3rd sub-problems (6 points) | Invoking physics principle in the 1st and 3rd subproblems: 2 points (1 point for each sub-problem) |
| | Applying physics principle correctly: 4 points (for each sub-problem, students received 2 points if they apply the physics principle correctly. One point was taken off if they made a mistake such as messing up with the mass(es) of the sphere(s) or mistakenly writing the kinetic energy as ½ $mv$ or $mv^2$, where $m$ stands for the mass, $v$ stands for the speed of the object(s) of interest ) |
| Conservation of Momentum in the 2nd sub-problem (4 points) | Invoking physics principle in the 2nd sub-problem: 1 point |
| | Applying physics principle correctly: 1 point |
| | Showed relevance of work to the final answer: 2 points |

    Students' performance in different intervention groups was later compared to that in the comparison group. Moreover, in order to examine the effects of interventions on students with different expertise and to evaluate whether the interventions were more successful in helping students at a particular level of expertise, we further classify the students in each course as top, middle or bottom based on their scores on the final exam. Students in the whole course (not distinguished between different recitation classrooms) were first ranked by their scores on the final exam. About 1/3 of the students were assigned to the top, middle, and bottom groups, respectively. The number of students in each case is shown in Table 4. As noted earlier, there was no significant difference between any of the groups in each course in terms of students' force concept inventory (FCI) score administered at the beginning of the semester. In order to take into account the possible difference between different recitation classes which may develop as the semester progresses, the overall performance of each intervention group is represented by an unweighted mean of students' performance from the three different levels of expertise[1]. We also compared the students' performance in these algebra-based and calculus-based introductory physics courses with the performance of a group of first-year physics graduate students who were asked to solve the quiz problem on their own without any solved problem provided. The

---

[1] We have also examined the results when the overall performance of each intervention group is represented by a weighted mean of students'performance from three different levels of expertise (i.e., each average is weighted by the corresponding number of students in that group with a particular level of expertise). Our analysis shows that the results are the same regardless of whether the overall performance is represented by a weighted mean or an unweighted mean. For simplicity, in this paper we will only show the data for the performance with an unweighted mean.

performance of the graduate students can serve as a benchmark for how well the undergraduate students can achieve as an upper limit.

Table 4 The number of students in each group in the calculus-based course and algebra-based course, where "Comp" stands for "comparison" and "Int" stands for "intervention".

| Group | Calculus-based | | | | Algebra-based | | | |
| --- | --- | --- | --- | --- | --- | --- | --- | --- |
| | Comp | Int 1 | Int 2 | Int 3 | Comp | Int 1 | Int 2 | Int 3 |
| Top | 13 | 13 | 13 | 19 | 10 | 27 | 21 | 15 |
| Middle | 12 | 10 | 10 | 35 | 19 | 11 | 17 | 17 |
| Bottom | 9 | 14 | 12 | 20 | 17 | 8 | 24 | 16 |
| Total | 34 | 37 | 35 | 74 | 46 | 46 | 62 | 48 |

### D. Interviews

In order to obtain an in-depth account of introductory physics students' reasoning while they solved the problem and explore additional strategies that may help them, six students from other introductory physics classes who didn't participate in the quiz were recruited for one-on-one interviews. Three of the six students we interviewed were enrolled in an algebra-based introductory mechanics course at the time of the interview; the other three students were enrolled in two different calculus-based mechanics courses. The interviews were conducted in the middle of the semester, after all the relevant topics had been covered in the lectures. All the students recruited for the interviews had a midterm score which fell in the middle of their own introductory physics course, ranging from +6 to -15 points above or below the class averages (which fell between 70% and 76% for different courses).

During the interviews, students were asked to learn from the solved snowboard problem provided and solve the isomorphic putty problem given. Similar to the previously discussed in-class quiz situation, different students in different interviews received different kinds of interventions, which are listed in Table 5. The interviews focused not just on understanding the difficulties students had, but also on examining the additional scaffoldings that may be helpful for the students. Some of the interventions were the same as the interventions used in the quantitative data. Some of the interventions were new in the sense that a slight modification was made to the interventions used in the quantitative data to explore additional strategies to help students. For example, in the interviews with students E and F, a new problem (the "two-block problem" shown in Appendix C) that is isomorphic to the solved problem and the quiz problem was introduced. This two-block problem, which consists of two steps: an object going down, colliding, sticking and moving together with another object on the horizontal part of the track, was designed in light of the common student difficulties found in the quantitative data. We will discuss the purpose of this new problem in more details later.

The audio-recorded interviews were typically 0.5-1 hour long. They were carried out using a think-aloud protocol, which allowed the researchers to follow and record students' thinking process. Students were asked to perform the task (whether they were reading the solved problem or trying to solve the quiz problem) while thinking aloud; they were not disturbed during the task. All the questions were asked to the students after they were completely done with the problem solving to the best of their abilities. After students completed the quiz, the researcher would first ask clarification questions in order to understand what they did not make explicit earlier and what their difficulties were. Based on this understanding, the researcher then provided some additional supports (sometimes including the physics knowledge required) to the students in order to help them solve the quiz problem correctly if they had not done so. The researcher also outlined or even demonstrated part of the solutions to the students as needed. After helping students learn how to solve the quiz problem correctly, the researcher invited them to reflect on the learning process they just went through (for example, by asking explicitly what was the thing that helped them figure out how to solve the problem) and asked them to provide some suggestion from the student's perspective on how to improve students' performance on the quiz (transfer) problem. The goal of the students' reflection was to help us identify the possible helpful scaffoldings not only based upon what the researchers observed but also based upon students' reflection of their own learning.

**Table 5.** The interventions students received in the interview.

| Student A (calculus-based) | Intervention 3 |
|---|---|
| Student B (algebra-based) | Intervention 3 |
| Student C (calculus-based) | Intervention 2 |
| Student D (algebra-based) | Intervention 2 |
| Student E (algebra-based) | Two Quiz Problems (version 1) |
| Student F (calculus-based) | Two Quiz Problems (version 2) |
| * Two quiz problems (version 1): (1) The student first learned from the solved snowboard problem provided and then solved another problem about "two blocks colliding" with the solved problem in his hand. (This "two-block" problem can be found in  (2) The researcher asked clarification questions in order to understand what the student did not make explicit earlier and to better understand their difficulties. The researcher also discussed with the student how to solve the "two block problem" correctly. (3) The student was asked to take advantage of what he learned from the previous two problems to solve the putty problem.<br>* Two quiz problems (version 2): (1) The student first learned from the solved snowboard problem provided and then solved the two quiz problems (the two-block bridging problem and the putty problem) with the solved problem in his hand. (2) The researcher asked clarification questions in order to understand what the student did not make explicit earlier and to better understand their difficulties. The researcher also discussed with the student how to solve the "two block problem" correctly. (3) The student was asked to take advantage of what he learned from the previous two problems and attempted to solve the putty problem a second time. ||

## III.  RESULTS AND DISCUSSION

### A. Student performances on the quiz problem

Table 6 shows the different answer categories for the answers graduate students provided when they were asked to solve the quiz problem on their own without scaffolding. The frequencies of each type of answer are listed. Twenty three out of 26 graduate students were able to figure out the 3-step nature of the solution even though some of them erroneously used $\frac{1}{2}mv$ instead of $\frac{1}{2}mv^2$ to calculate the kinetic energy or made mistakes related to the masses on the two sides of the equation in the 3rd step. Two graduate students incorrectly claimed that the total mechanical energy was conserved throughout (including all the processes), forgetting about the fact that there was an inelastic collision involved in which some mechanical energy will be transformed into other forms of energy when two objects stick together. The principle of CM was not invoked in these two students' solutions. The 26 graduate students on average scored 9.2 out of 10 on the quiz problem when scored using the rubric shown in Table 3.

**Table 6.** Graduate students' answer categories and frequencies for the putty problem.

| Descriptions of Graduate Students' Answers (In the following equations, $m$ and $v$ stand for the mass and the speed of the object(s) of interest, respectively) | Number of students |
|---|---|
| Correct 3-step solution: $$m_A g h_o = \frac{1}{2} m_A v_A^2 \Rightarrow v_A = \sqrt{2gh_o}$$ $$m_A v_A = (m_A + m_B) v_{A+B} \Rightarrow v_{A+B} = \frac{\sqrt{2gh_o}}{2}$$ $$\frac{1}{2}(m_A + m_B) v_{A+B}^2 = (m_A + m_B) g h_f \Rightarrow h_f = \frac{1}{4} h_o$$ | 20 |
| Correct except that in the 3rd step, the student used $mgh = \frac{1}{2}mv$ | 1 |
| Correct except that in both the 1st and 3rd step, the student used $mgh = \frac{1}{2}mv$ | 1 |
| Correct except that in the 3rd step, the masses on the two sides of the equation are not consistent $mgh_f = \frac{1}{2}(2m)v_{A+B}^2$ | 1 |
| $m_A g h_o = \frac{1}{2}mv^2 = (m+m)gh_f$ | 1 |
| $mgh_o = 2mgh_f$ | 1 |
| Both $m_A g h_o = \frac{1}{2} m_A v^2 = (m_A + m_B) g h_f$ and 3-step solution (but in the 3rd step the student used $\frac{1}{2}(m_A + m_B) v_{A+B} = (m_A + m_B) g h_f$) | 1 |

As for the introductory students, Table 7 and Table 8 present students' average scores on the quiz (transfer) problem in the calculus-based and algebra-based courses, respectively. Due to the instructor's time constraint in the recitation classes, the allotted time for students in intervention group 2 to try the quiz problem on their own before learning from the solved problem was slightly less than the time given to those in the comparison group. Therefore, instead of examining how intervention 2 students' pre-scaffolding performance compares to that of the comparison group, in these tables we only focus on the performance of students in intervention group 2 AFTER the scaffolding support. As shown in Table 7 and Table 8, the average scores of the comparison group students in the two courses indicate that many students had great difficulty with the putty problem. Even though students in the three intervention groups received the solved problem and other scaffolding supports to help them solve the quiz (transfer) problem, their performance didn't show great improvement. In the calculus-based course, the comparison group students who solved the quiz problem on their own received an average score of 6.3 out of 10. The average scores of the three intervention groups were similar. Analysis of variance (ANOVA) indicates that none of the intervention groups in the calculus-based course show a statistically different performance from that of the comparison group. In the algebra-based course, even though the scores went up significantly ($p < 0.05$) from 2.5 (in the comparison group) to 4.4, 5.4, and 5.2 in the three intervention groups, respectively, there is still much room for improvement. It turns out that this problem was challenging for the calculus-based students and even more difficult for the algebra-based students. The p-values, which compares the performance of the comparison group students and various intervention group students, are listed in Table 9.

**Table 7.** Students' average scores out of 10 on the quiz (transfer) problem in the calculus-based course.

|  | Comparison | Intervention 1 | Intervention 2 | Intervention 3 |
|---|---|---|---|---|
| Top | 8.2 (3.1) | 9.2 (2.2) | 8.4 (2.4) | 8.2 (2.9) |
| Middle | 6.8 (2.9) | 6.1 (3.4) | 8.4 (2.1) | 6.9 (3.1) |
| Bottom | 3.9 (3.2) | 3.8 (3.2) | 5.2 (3.1) | 5.4 (3.3) |
| Average | 6.3 | 6.4 | 7.3 | 6.8 |

Note: The standard deviation in each case is shown in parentheses. The performance of the whole group taken together is represented by an unweighted mean of students' average scores from the top, middle and bottom categories.

**Table 8.** Students' average scores out of 10 on the quiz problem in the algebra-based course.

|  | Comparison | Intervention 1 | Intervention 2 | Intervention 3 |
|---|---|---|---|---|
| Top | 3.8 (2.9) | 5.3 (3.2) | 7.3 (3.0) | 6.2 (3.1) |
| Middle | 1.9 (2.5) | 3.3 (3.0) | 4.2 (2.3) | 5.3 (1.8) |
| Bottom | 1.9 (1.7) | 4.5 (3.2) | 4.6 (2.7) | 4.2 (1.6) |
| Average | 2.5 | 4.4 | 5.4 | 5.2 |

Note: The standard deviation in each case is shown in parentheses. The performance of the whole group taken together is represented by an unweighted mean of students' average scores from the top, middle and bottom categories.

**Table 9.** The p values for the comparison of students' performance between the comparison group (Comp) and different intervention (Int) groups in the calculus-based and algebra-based courses.

|  | Comp v.s. Int 1 | Comp v.s. Int 2 | Comp v.s. Int 3 | Int 1 v.s. Int 2 | Int 1 vs. Int 3 | Int 2 v.s. Int 3 |
|---|---|---|---|---|---|---|
| Calculus | 0.994 | 0.772 | 0.976 | 0.597 | 0.882 | 0.894 |
| Algebra | 0.001 | 0.000 | 0.000 | 0.561 | 0.810 | 0.984 |

Table 10 presents intervention 1 students' score on the isomorphic snowboard problem reproduced immediately after browsing over and returning its solution. Table 10 shows that the calculus-based and algebra-based students achieved an average score of 9.6 and 8.3 out of 10, respectively, on the snowboard problem. These results suggest that both algebra-based and calculus-based students were generally good at reproducing the solved problem they just learned from[2]. However, as shown in Table 7 and Table 8, the high score they achieved on the solved problem reproduced didn't automatically imply that they had the ability to transfer their learning to the isomorphic quiz problem. On average, the quiz (transfer) problem scores by the same group of students dropped by 3.2 and 3.9 points out of 10 in the calculus-based and algebra-based courses, respectively.

**Table 10.** Average scores out of 10 on the snowboard problem (solved problem) and the putty problem (quiz problem) for intervention 1 in the algebra-based and calculus-based courses.

|  | Solved Problem | | Quiz Problem | |
|---|---|---|---|---|
|  | Calculus | Algebra | Calculus | Algebra |
| Top | 9.9 | 8.8 | 9.2 | 5.3 |
| Middle | 9.9 | 6.8 | 6.1 | 3.3 |
| Bottom | 8.9 | 9.4 | 3.8 | 4.5 |
| All | 9.6 | 8.3 | 6.4 | 4.4 |

Comparing the results of our study with that of the previous study [46] indicates that it is very difficult for students to extend what they learned from a two-step problem to solve a three-step problem as compared to simply going from a two-step problem to another two-step problem. In the previous study [46] in which students were asked to take advantage of what they learned from the solved roller-coaster problem to solve the new tire swing problem, we found that students performed reasonably well in transferring what they learned from the 2-step solved problem to solve the 2-step quiz problem. For example, in that study, the calculus-based students in the intervention group 2 received an average score of 9.1 out of 10 on the quiz problem, which is not only significantly better than the average score (6.8 out of 10) of the comparison group students, who didn't receive any scaffolding support, but also higher than the benchmark (8.4) set by the graduate students.[46] However, in this study, even though the same scaffolding support was implemented, students in the intervention group 2 didn't significantly outperform the comparison group. In fact, none of the intervention group in the calculus-based course was significantly better than the comparison group, nor is the average score in any of the groups comparable to the benchmark set by the graduate students. Although there is different content knowledge involved in these two studies (i.e., the Newton's 2$^{nd}$ Law in the non-equilibrium situation with a centripetal acceleration involved versus the principle of conservation of momentum), we do not believe the content of conservation of momentum is the major reason that makes the transfer in this study extremely challenging. In fact, since the problems used in the previous study require the application of Newton's 2$^{nd}$ Law in the non-equilibrium situation with a centripetal acceleration involved, they may be considered more difficult than the problems involving the principle of conservation of momentum. For example, when the rollercoaster problem, tire swing problem, and snowboard problem were used in another study [66,67], it was found that if the target problem was also a 2-step problem, students who self-diagnosed their own mistakes in the snowboard problem were capable of transferring their learning to solve an isomorphic 2-step problem. Comparably, when the same self-diagnosis task was given, the transfer from the 2-step rollercoaster problem to the 2-step tire swing problem was found to be more difficult. In view of these previous studies, we believe that the great difficulty students had in our study transferring from the snowboard problem to the putty problem lies in the fact that going from a 2 step problem to a 3 step problem was very challenging for students. Since the putty problem has an additional step compared to the snowboard problem, students cannot correctly solve the putty problem by blindly replicating the procedures shown in the solution to the snowboard problem. Rather, they have to be able to decompose the problem appropriately into temporally separated sub-problems based on the appropriate physics principles. In addition, they have to be able to figure out how the different sub-problems should be connected. If students don't have a holistic picture of the 3-step structure of the problem solution, they are likely to make mistakes.

---

[2] Although the middle group in the algebra-based course has a lower average score on the solved problem, examination of students' performance in details indicates that this score may have been affected by few students with extreme scores since there is a total of only 11 students in this group. Our data show that two of the students in this group received a score of 0 and 3 on the solved problem, respectively, while all the other students received a score of 5 or more. If those 2 students were not included, the average score on the solved problem would be 8 for the middle group, and 8.7 for the "all" category).

Examining students' written work from the two introductory physics courses in detail and exploring students' problem solving performance during the think aloud interviews further confirm our findings that it was very challenging for students to figure out the 3 step nature of the quiz (transfer) problem. Table 11 presents the common mistakes that the introductory students made. Table 12 presents the corresponding percentage of students who made each mistake listed in Table 11. The percentage of students who display good understanding of the three step nature of the solution structure (even though they may have made some mistakes in the application details such as using $1/2mv$ for kinetic energy or using the wrong mass on one side of the CME equation similar to those presented in Table 6) is also listed for comparison. As shown in Table 12, only 44~50% of the calculus based students and less than 25% of the algebra-based students were able to figure out the 3 step nature of the quiz problem. Examination of introductory students' work indicates that forgetting to invoke the principle of CM is one of the most common mistakes they made when no scaffolding was provided, especially in the algebra-based course. Twenty four percent of the calculus-based control group students and fifty percent of the algebra-based control group students made this mistake. About half of these students simply related the initial potential energy of putty A (when it is raised to the initial height $h_o$) to the final potential energy of putty A and B (when both of them reach the maximum height $h_f$) and came up with an expression $m_A gh_o= (m_A+m_B)gh_f$ without considering the process in between.

Some other students took into account the intermediate process but still came up with a similar answer $m_A g h_o = 1/2 \, mv^2 = (m_A + m_B) g h_f$. (Depending on the student, $m$ and $v$ here could stand for the mass and the speed of putty A right before the collision, or the mass and the speed of both putties together right after the collision.) Examination of students' work in the intervention groups indicates that even for students who recognize that the CM principle is applicable to the collision process after learning from the solved problem, they might not necessarily understand the implication of applying this principle in their solutions. For example, some of the intervention group students successfully found that the speed of two putty spheres together immediately after the collision would be half of the speed of putty A right before the collision by using CM principle, but they just left it aside after that and did not make use of it later. They resorted to other ideas (e.g., $m_A g h_o = (m_A + m_B) g h_f$) to come up with the final answer. An example of this type of student's work was shown in Figure 1. The think aloud interviews suggest that one reason that the students invoked the CM principle initially but didn't make use of it later in their solution may be due to the fact that students in general didn't systematically come up with a plan for solving the problem before implementing the plan. When solving the quiz problem after browsing over the solved problem, many students first wrote down the principle they "believed" should be used (because the same principle was shown in the given solved problem) and then tried to plug in some variables from the new situation in order to solve for the target variable. However, because the solution to the quiz problem had many differences compared to the solved problem, such strategies didn't get them too far. After writing down an equation using the CM principle, these students didn't know what to do with that equation and didn't know how to connect it to the target variable. They just left it aside and started working on a new equation (e.g., the conservation of mechanical energy equation) without coming back and relating it to their original work with the momentum principle.

**Table 11. Common student approaches to and/or mistakes in the quiz problem.**

| Category | Description of introductory students' approaches and/or mistakes in each category. Examples of students' work in each category are also provided. |
|---|---|
| No major structural mistake | Students were able to figure out the 3 step nature of the putty problem (even though they might have made some mistakes in the application details such as erroneously using $\frac{1}{2}mv$ to calculate the kinetic energy or making mistakes related to the masses on the two sides of the equation).<br><br>*Example 1: Correct 3-step solution:*<br>$$m_A g h_o = \frac{1}{2} m_A v_A^2 \Rightarrow v_A = \sqrt{2gh_o}$$<br>$$m_A v_A = (m_A + m_B) v_{A+B} \Rightarrow v_{A+B} = \frac{\sqrt{2gh_o}}{2}$$<br>$$\frac{1}{2}(m_A + m_B) v_{A+B}^2 = (m_A + m_B) g h_f \Rightarrow h_f = \frac{1}{4} h_o$$<br><br>*Example 2:*<br><br>*Correct except that in the 3$^{rd}$ step, the student wrote $mgh_f = \frac{1}{2}(2m)v_{A+B}^2$ (the masses on the two sides of the equation are not consistent)*<br><br>Note: In order to be placed in this category, students cannot make any of the mistakes described in the following categories. |
| CM_A:<br><br>No CM | Didn't invoke CM principle in the solution |
| CM_B:<br><br>Wrong CM | Invoked CM principle in situations that are clearly incorrect (e.g., from the very beginning to the very end)<br><br>Example: $m_A v_A + m_B v_B = (m_A + m_B) v_{A+B}$<br>$$0 = (m_A + m_B) v_{A+B}$$ |
| CME_1:<br>CME across collision | Using conservation of mechanical energy in processes that involves the inelastic collision in between (regardless of whether the student invoked the CM principle or not)<br><br>*Example 1:* $m_A g h_o = (m_A + m_B) g h_f$ |
| | *Example 2:* $m_A g h_o = \frac{1}{2}(m_A+m_B)v_{AB}^2$

*Example 3:* $\frac{1}{2} m_A v_A^2 = (m_A+m_B)g h_f$

*Example 4:* $m_A g h_o = m_A g h_f$ |
| CME_2: Summing up KE and PE from different positions | (a) Combining several processes into one and summing up the potential energy in one situation with the kinetic energy at another situation, and (b) Using conservation of mechanical energy in processes that involves the inelastic collision in between

(regardless of whether the student invoked the CM principle or not)

*Example 1:* $m_A g h_o + \frac{1}{2} m_A v_A^2 = \frac{1}{2}(m_A+m_B)v_f^2 + (m_A+m_B)g h_f$

*Example 2:* $m g h_o + \frac{1}{2} m v_A^2 = (2m)g h_f$

*Example 3:* $m_A g h_o = \frac{1}{2}(m_A+m_B)v_f^2 + (m_A+m_B)g h_f$

*Example 4:* $m g h_o + \frac{1}{2} m v_{A+B}^2 = m g h_f$

Note : As discussed in Figure 2, since most students didn't explicate what their variables (especially the various "$v$"s) in their equations mean, it is possible that a solution like the examples shown here would finally lead to a solution similar to those presented in the previous category (CME_1) if the speed(s) are set to be zero. However, unless an explicit indication that the velocity(ies) are zero is found in their work, all the velocities are assumed to be nonzero, and students' work is categorized in this group, not in group CME_1. |
| CME_3: Resembling the solved problem | Having solutions very similar to that of the solved problem (i.e., having one CM principle in which the students find that $m_A v_A = (m_A + m_B)v_{A+B}$ and one CME principle in which the students use $\frac{1}{2}m v_{A+B}^2 = mgh$), but failing to figure out the exact structure to solve the problem.

*Example 1:* Correctly mapping the solution of the snowboard problem to the last two sub-problems of the putty problem (i.e. using h=hf), but not knowing what to do with the 1$^{st}$ sub-problem that was not included in the solved problem and just leaving it unused.

*Example 2:* after finding $\frac{m_A v_A}{2m_A} = v_{A+B}$, incorrectly using $v_{A+B} = \sqrt{2gh} = \sqrt{2g(h_o - h_f)}$ and coming up with a final answer
$$h_f = h_o - \frac{1}{g}\left(\frac{m_A v_A}{2m_A}\right)^2$$ |

| | |
|---|---|
| CME_4: V the same | Velocity of the sphere is the same right before and right after the collision<br><br>*Example:* $\quad m_A g h_o = \frac{1}{2} m_A v^2 \Rightarrow v^2 = 2g h_o$<br>$\quad\quad\quad\quad\quad \frac{1}{2}(m_A+m_B)v^2 = (m_A+m_B)g h_f$<br>$\quad\quad\quad\quad\quad h_f = v^2/2g = h_o$ |
| CME_5: other structural mistake | Other student mistakes (including but not limited to the following cases):<br><br>*Example1 : using 1-D kinematics equations instead of CME principle*<br><br>*Example 2: Only writing down some mechanical energy terms/variables separately without writing down any equation to connect them together*<br><br>*Example 3: directly answering that hf=ho/2 without any solution process provided (or the written solution process doesn't lead to the final answer provided)*<br><br>*Example 4: Only some scribbles without a clear solution process* |

**Table 12. Percentage of students in each group who adopted the different approaches or made the mistakes listed in Table 11.**

|  | Calculus-based course | | | | Algebra-based course | | | |
|---|---|---|---|---|---|---|---|---|
|  | comp | Int 1 | Int 2 | Int 3 | Comp | Int 1 | Int 2 | Int 3 |
| No major structural mistake | 44.1 | 48.6 | 48.6 | 50.0 | 6.5 | 15.2 | 24.4 | 10.4 |
| CM_A: No CM | 23.5 | 29.7 | 11.4 | 16.2 | 50.0 | 28.3 | 14.5 | 10.4 |
| CM_B: Incorrect CM | 14.7 | 2.7 | 0.0 | 8.1 | 6.5 | 2.2 | 0.0 | 0.0 |
| CME_1: CME across collision | 26.5 | 21.6 | 8.6 | 20.3 | 28.3 | 21.7 | 16.1 | 20.8 |
| CME_2: Summing up KE and PE from different positions | 14.7 | 16.2 | 11.4 | 14.9 | 28.3 | 34.8 | 24.2 | 22.9 |
| CME_3: Resembling the solved problem | 0.0 | 2.7 | 17.1 | 6.8 | 2.2 | 10.9 | 17.7 | 29.2 |
| CME_4: V the same | 5.9 | 0.0 | 2.9 | 2.7 | 0.0 | 0.0 | 0.0 | 0.0 |
| CME_5: other structural mistakes | 8.8 | 10.8 | 11.4 | 5.4 | 34.8 | 17.4 | 17.7 | 16.7 |

Note: A single student may make both a mistake related to CM principle and a mistake related to CME principle. (For example, a student who had a solution $m_A g h_o = (m_A + m_B) g h_f$ would be categorized as having made 2 mistakes: "CM_A: no CM" and "CME_1: CME across collision".) Therefore, the numbers in each column would sum up to 100% only if neither category A or category B are counted for a mistake.

$$mv_A = 2mv_{AB}$$
$$v_{AB} = \frac{1}{2}v_A$$
$$KE_{fA} = KE_{iAB}$$
$$PE_{iA} = PE_{fAB}$$
$$mgh_o = 2mgh_f$$
$$h_f = \frac{1}{2}h_o$$

**Figure 1.** An example of an introductory student's answer to the putty (transfer) problem. Even though the student invoked the CM principle, he didn't use this principle to find the final answer. (Fig 1)

Another common mistake that students made was invoking the CME principle inappropriately in situations that involve the inelastic collision. In Table 11 and Table 12, all students placed in category CME_1 ("CME across collision") and category CME_2 ("Summing up KE and PE from different positions") had this mistake. As shown in Table 12, this mistake was prevalent in both the control and the intervention groups, suggesting that even though students recognized the similarity between the isomorphic problems in terms of the principles involved and invoked the relevant principles, many of them didn't understand the situations in which each principle should be applicable. The mistake of treating the mechanical energy as conserved throughout the whole process and coming up with a final answer $h_f = \frac{M_a}{M_a+M_b}h_o = \frac{1}{2}h_o$ (categorized in category CME_1:"CME across collision") is one common type of students' work that involves the mistake of invoking the CE principle in inappropriate situations. Figure 2 presents another example that represents another common type of students' incorrect solution. In Figure 2, there were 2 steps involved in the solution; the first step involved the CM principle and the other involved the CME principle. Why the student applied these two principles in the manner he did, however, is not clear. One way of interpreting the student's solution to the transfer problem is to assume that $v_A$ and $v_f$ stand for the speed of putty A right before the collision and the speed for both putty A and B together immediately after the collision. If this assumption is correct, the student would have applied the principle of CM correctly to the collision process but made a mistake with the CME part because the student erroneously combined the initial potential energy of putty A (when it was released) with the kinetic energy at a later instance (when putty A reached the bottom) and set it equal to the kinetic energy of putties A+B together right after the collision plus the final potential energy of putties A+B when they reached the maximum height. The mistake of summing up potential energy and kinetic energy from different instances on one side of the CME equation indicates that the student didn't fully understand the meaning of the CME principle and he didn't know how to apply it correctly. Figure 3 is another example of a student's solution which shows a similar mistake of mixing up several processes into one and applying the CME principle to an incorrect situation. Similar mistakes were classified in the "CME_2: summing up KE and PE from different positions" group in Table 11 and Table 12.

$$\Delta \vec{P} = 0 \quad \vec{P}_f = \vec{P}_i \quad 2mV_f = mV_A + m\cancel{V_B} \quad \begin{matrix} V_B = 0 \\ V_f = \dfrac{1}{2}V_A \end{matrix}$$

$$\Delta E_{mec} = \frac{1}{2}mV_A^2 + mgh_o = \frac{1}{2}2mV_f^2 + 2mgh_f$$

$$h_f = \frac{\frac{1}{2}mV_A^2 + mgh_o - mV_f^2}{2mg}$$

$$h_f = \frac{\frac{1}{2}V_A^2 + gh_o - \left(\frac{1}{2}V_A\right)^2}{2g}$$

$$h_f = \frac{\frac{1}{2}V_A^2 + gh_o - \frac{1}{4}V_A^2}{2g}$$

$$h_f = \frac{gh_o - \frac{1}{2}V_A^2}{2g}$$

**Figure 2.** An example of a student's incorrect answer. The situations in which the CM and CME principles were applied are not clear.

$$PE_1 = m_A gh_o \quad PE_f = m_{A+B} gh_f$$

$$m_A gh_o + \frac{1}{2} m_{A+B} v_{A+B}^2 = m_{A+B} gh_f$$

$$P_1 = P_f$$
$$m_A v_A + m_B \cancel{v_B} = (m_A + m_B) v_{A+B}$$
$$v_{A+B} = \frac{m_A v_A}{m_A + m_B}$$

$$m_A gh_o + \frac{1}{2} m_{A+B} \left(\frac{m_A v_A}{m_A + m_B}\right) = m_{A+B} gh_f$$

$$h_f = \frac{m_A gh_o + \frac{1}{2} m_{A+B} \left(\frac{m_A v_A}{m_A + m_B}\right)}{m_{A+B} \cdot g}$$

**Figure 3.** Another example of a student's solution in which the student mixed up several processes into one and applied the CME principle to an incorrect situation. Similar mistakes were classified in the "CME_2: summing up KE and PE from different positions" category in Table 11.

Another possible way to interpret the student's solution in Figure 2 is to postulate that the student realized he should only combine the potential energy and kinetic energy of a system at the same moment on one side of the equation. In this case, $v_A$ and $v_f$ would stand for the speed of putty A when it was released and the speed of putty AB together when they momentarily reached the maximum height $h_f$, which would mean that $v_A$ and $v_f$ should both be zero. The student may then be thinking about the mechanical energy being conserved during the whole process, which can be reduced to the previously described common mistake of setting $m_Agh_o= (m_A+m_B)gh_f$. The student, however, would have invoked the CM principle in an incorrect situation. Figure 4 shows an example of the solution of another student who explicitly claimed that the initial momentum of the system equals the final momentum of the system where all the speeds involved were zero. This mistake suggests that the student didn't realize that the CM principle is applicable only during the collision process and not during the entire process. The percentages of students who explicitly applied the CM principle in an incorrect situation are listed in Table 12 (presented in group "CM_B: wrong CM"). We note that when binning student mistakes into catergories in Table 12, when there is an ambiguity in interpreting the meaning of the velocity(ies), all the velocities are assumed to be nonzero unless an explicit indication that the velocity(ies) are zero can be found in their work.

$$m_A v_{Ai} + m_B v_{Bi} = v_{fA+B}(m_A + m_B)$$
$$m_A(0) + m_B(0) = 0(m_A + m_B)$$
$$0 = 0$$

$$\cancel{KE_i} + PE_i = \cancel{KE_f} + PE_f$$
$$PE_i = PE_f$$
$$m_A g h_i + m_B g h_i = (m_{A+B})(g)(h) \qquad m_A = m_B \therefore m_A + m_B = 2m$$
$$(m)_A \cancel{(g)}(h_i) = (2m)\cancel{(g)}(h_f)$$
$$\boxed{\tfrac{1}{2} h_i = h_f}$$

**Figure 4.** Example solution by a student who applied the CM principle to an incorrect situation.

In addition to the mistakes previously mentioned, other common student mistakes include having a solution that resembles the solution to the solved problem but failing to figure out the exact structure to solve the problem, treating the speed of the sphere right after the collision the same as the speed of the sphere right before the collision, etc. Examples of these mistakes are presented in Table 11. We note that Table 11 and Table 12 and present the distribution of common student answers/mistakes as shown on the quiz. In addition to those tabulated mistakes, there could be other underlying difficulties that students had, which may have led them to arrive at the answers shown in Tables 11 and 12. In the following paragraphs, we will discuss these possible underlying difficulties. Findings from the interviews that shed light on these issues will be presented.

First, we found that one important reason that the students couldn't figure out the 3 step nature of the problem may have to do with the fact that the students may not have a clear picture in mind about what was going on in the quiz problem and they may not necessarily think through the problem in sufficient depth. For example, in the interviews, we found that most students failed to describe the meaning of each variable in a precise way. When the researcher asked students to explicate what each variable, especially the various "$v$"s mean in the interviews (for example, in the equation $m_A v_A + \cancel{m_B v_B} = (m_A + m_B)v_{A+B}$), they often answered that $v_A$ was the velocity "before collision", and $v_{A+B}$ was the velocity "after collision". It was not easy for them to articulate by themselves which velocity before collision they were talking about (e.g., whether it was the velocity of sphere A "right before the collision", or the velocity of sphere A "at the very beginning when it was released"). It was possible that the students initially didn't recognize that in both "before the collision" and "after the collision", there were processes involved in which the speed of the sphere(s) varied with the height. Sometimes it took the researcher some effort to explain to the students that "velocity before/after the collision" could mean many different things since at different heights the velocities were different. As suggested by one student during the reflection, the idea of "snapshots of the putty at different points" was very helpful in solving the problem. However, the student was able to articulate this idea only after the discussion with the researcher during the interview.

Moreover, it is possible that the concept of the "infinitesimal" time before and after the collision involved in the CM principle was very challenging for the students. If students didn't realize that momentum of the two putty system was conserved only immediately before and after the collision, they were likely to apply the CM principle in incorrect situation as presented in Table 11, category CM_B. In all the interviews, students were asked explicitly to identify the situation in which their CM principle was applied by describing what their variables $v_i$ and $v_f$ in the CM equation refer to. They were sometimes given the following choices to help them articulate their answers: their CM principle was applied (1) from the very beginning to the very end (2) from right before the collision to right after the collision (3) somewhere in between (1) and (2). During the interview, one student responded in the following manner:

*Student B : Wouldn't they...Wouldn't the total momentum for the system be the same throughout? Or would [it] not be?*
*Researcher: What do you think?*
*Student B: I think it would be.*
*Researcher: OK. So…why?*
*Student B: Um…just because from what I heard for the conservation of momentum, from what I've been told about, momentum is conserved throughout…Uh…yea, no matter what time it is, whether it's t [equals] zero or t [equals] infinity, the momentum should be conserved throughout.*
*Researcher: OK. Do you remember…when your professor taught you about conservation of momentum, did he or she say when should that principle be applicable?*
*Student B: Um…I believe…it may have been not applicable for inelastic collision but I don't really remember. And I believe it [the solution to the snowboard problem] says that this was inelastic… Um…I believe you would use it [conservation of momentum] more in elastic collision than you would in inelastic collision.*

The dialogue above suggests that when the student exploited the CM principle to solve the quiz problem, he didn't carefully examine the applicability of the CM principle by considering whether or not there was an external force acting on the system, and he didn't notice that the momentum is only conserved from immediately before to immediately after the collision. In fact, the interaction between the researcher and the student suggests that it is very likely that the student didn't know how to correctly examine the applicability of the CM principle. When the researcher asked him about the situations in which the CM principle could be applicable, the student discussed the applicability incorrectly based upon whether the collision was elastic or inelastic. Findings from our study suggest that more scaffolding supports may be required to help students deal with this issue and acquire a better understanding of the situation in which CM principle can be applied.

Another important student difficulty that hinders students from solving the putty problem correctly may be that the students may not necessarily understand the role of inelastic collision for mechanical energy conservation. For example, for students who made the mistakes CME_1 and CME_2 in Table 11 and Table 12, it is likely that they did not realize that during the inelastic collision, some mechanical energy would be transformed into heat or other forms of energy and therefore the mechanical energy of the system was not conserved. During the interviews, when the students were asked about the difference between elastic and inelastic collisions, most students discussed the difference between two collisions by some surface feature such as (i) whether the object keeps its original shape and/or (ii) whether the two objects move apart or become one after the collision, instead of mentioning the definition of an elastic (inelastic) collision as a collision in which mechanical energy is (is not) conserved. When the researcher asked them to discuss further the specific principles that were applicable or not applicable in the elastic and inelastic collisions (e.g, by directly guiding them to consider the principles of CME and CM), many of them didn't know the correct answer and was unable to make up his mind. Quotes from two students are listed below. This finding suggests that even if students could recognize the inelastic collision process involved in the putty problem, they did not necessarily understand that the mechanical energy was not conserved throughout the whole process. More help may be required to help students in solving the putty problem correctly.

*Researcher: Do you remember if this collision is an elastic or inelastic collision?*
*Student C: Uh. Yea. Elastic means that they come together and then they can retain their shape, right? like... when they hit together, then when they come ap[apart]...they can come apart, and so be the same shape as they were when they came into collision. And then... and... inelastic forces they hit and they are one object.*
*Researcher: do you mean they stick...*
*Student C: Yea they stick together and they stick together throughout the system.*
*Researcher: Ok. So do you still remember which principle is applicable during the collision process and which is not? I mean, is there any difference between these two kinds of collision?*
*Student C: Uh...I mean... there is. I just...I just don't know.*

*Researcher: How about... let's think about the momentum?*
*Student C: Yea, it would be the difference between...like.... conserves momentum and [does] not conserves momentum? Would that be it?*
*Researcher: What do you think?*
*Student C: I... I... I actually don't know.*
*Researcher: Ok. So…. momentum is conserved in both kinds of collisions.*
*Student C: Oh, really?*
*Researcher: Yea.*
*Student C: Ok.*
*Researcher: So do you still remember anything about energy?*
*Student C: I do. I... I remember like... kinetic energy equ[equation]... uh... conservation of energy equation, just 'cause that was what we were taught a lot about. And then, we*

> *are taught a little bit about momentum, but… it shoots off very quickly. And I haven't taken physics before, so it's all new to me. So…*

*Researcher: Yea, I understand that. Don't feel bad… I mean… we know the [CM] principle is very difficult for lots of students.*

*Student C: OK.*

*Researcher: So…um… you are not quite sure whether energy is conserved during…*

*Student C: between elastic and inelastic?*

*Researcher: Uh huh. Do you remember anything about that?*

*Student C: Uh...energy is conserved in a…inelastic collision? Or? Uh….Let me think about this.*

*Researcher: Yea, take your time.*

*Student C: Uh….Uh….I guess energy is… conserved in inelastic collision? Yeah, I think so.*

*Researcher: So is energy conserved in an elastic collision?*

*Student C: Uh…is energy conserved in an elastic collision…*

*[silence]*

*Student C: Uh...I attempt to say no. But at the same time I don't think so. I don't think it's yes. I'm going to go with no because once you hit it, like you… like for this object, when the one ball hits the other ball, and if it were inelastic, they come together, and all the energies is gonna fall in with them. Whereas if the one hits, this one is still on the, like going up a little bit...but this one is also going up, so...uh…I guess energy is … I'm going to go with energy is conserved in both.*

*Student F: There's one where you can use conservation of momentum and energy. And there's another one where you can use one of those. But I don't remember which one [elastic or inelastic] it goes to and which one [CM or CME] works.*

Not being able to figure out the 3 step nature of the quiz problem and mistakenly using the CME principle only once (in processes involving the inelastic collision such as those shown in groups CME_1 and CME_2 in Table 11) was the major difficulties many students had. Therefore, one may expect that intervention 3, which involves an additional hint about "using the conservation of mechanical energy twice", may be of great help to the students. However, the quantitative data obtained from the two introductory physics courses (please see Table 7, Table 8, and Table 12 for example) suggests that this intervention didn't help students significantly. The same phenomenon was observed in the interview as well. Figure 5 presents the solution of a student who was subjected to intervention 3 during the think aloud interview. Initially, he didn't appear to pay much attention to the additional hint about "using the conservation of mechanical energy twice" and in his work, CME was used only once. After he had tried the putty problem on his own to the best of his knowledge (shown in Figure 5), the researcher reminded him that the snowboard problem was a 2-part problem and the putty problem was a 3-part problem in which we had to use the CME principle twice. However, this reminder didn't help, either. He still thought that the 1$^{st}$ step in solving the putty problem was to use the CM principle similar to what he did in his earlier work. He felt that the putty problem must be solved by using CM first, followed by the use of CME twice. He incorrectly interpreted that it was the last CME part which would lead him to express $h_f$ in terms of $h_o$ that he missed in his original work, but he struggled greatly and didn't know what to do with it. This interview suggests that some students might interpret the instruction of "using CME twice" in a different way than was intended, which could be one possible reason why providing the additional instruction that might be considered an additional hint by the experts in intervention group 3 didn't work as intended for the students.

Overall, the data from the two introductory courses as well as the interviews suggest that in this analogical problem solving and transfer activity, although students may be able to recognize the similarity between the isomorphic problems in terms of the principles involved, the solution to the snowboard problem doesn't necessarily help them figure out the structure of the three part putty problem and apply the principles correctly. In general, our research suggests that more algebra-based students benefited from the solved problem provided than the calculus-based students in the sense that the average scores in the intervention groups are significantly better than that in the comparison group. As shown in Table 12, many algebra-based control group students simply had no clue about how to solve the putty (transfer) problem when no scaffolding was provided. Some of them invoked the 1-D kinematics equations and weren't able to go far after that. Without additional support, some of them simply wrote down some potential-energy-like or kinetic-energy-like terms separately without writing any equation. (These mistakes were all classified in category CME_5 "other structural mistake" in Table 12.) The fact that with the solved problem provided, most of the algebra-based students were able to invoke either one or both of the correct principles (which they were not able to do when asked to solve the problem on their own) is the main reason why all three intervention groups performed significantly better than the comparison group in the algebra-based courses. However, the not so high absolute scores (around 44% to 54%) after the scaffolding also reflects the fact that the algebra-based students weren't necessarily able to apply the relevant principle correctly. For the calculus-based course, on the other hand, students typically were able to invoke the relevant principle(s) even without being provided the solved problem. The main difficulty for the calculus-based students therefore lay in how to proceduralize these principles in an appropriate manner. As we found in this study, our current scaffolding support provided (i.e., the solution to the snowboard problem together with different interventions implemented) was not effective in this regard; the difficulty of applying the principles correctly still remained even after they received the scaffolding.

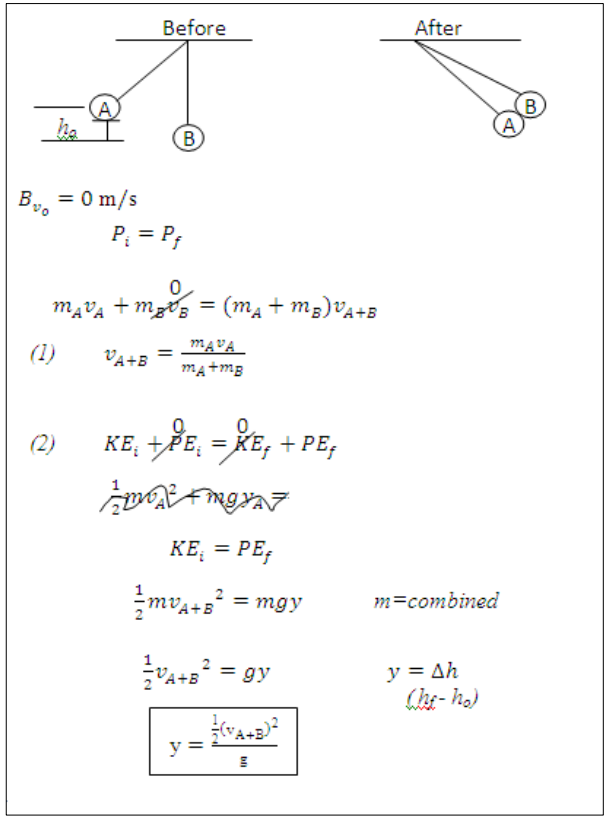

**Figure 5.** Students A's answer to the putty problem.

**B. Effect of additional scaffolding (beyond those used in the two introductory physics courses) as suggested by the interviews**

As mentioned earlier, one important goal of the interviews was to examine the possible additional scaffoldings (i.e., scaffoldings beyond those used in the two introductory physics courses) that may help the students solve the quiz (transfer) problem correctly. The different scaffoldings that were tried in the interviews to help students solve the problem successfully included:
<i> Directing students' attention to the fact that energy is not conserved in an inelastic collision.
<ii> Directing students' attention to the fact that the CM principle is only valid right before to right after the collision
< iii> Telling students explicitly that the problem can be decomposed into 3 parts.
<iv> Helping students learn how to solve a simpler 2-step problem (the bridging two-block problem) first.
As noted earlier, depending on the prior knowledge and difficulties students had, different students received different levels of additional scaffolding in the interviews to help them solve the quiz problem correctly. A detailed list of the additional scaffoldings that each interviewed student received as well as the effects of these scaffoldings can be found in Appendix D. (Student B was not included in Appendix D because there was not enough time in the interview to provide him with additional scaffoldings to help him solve the putty problem correctly.)

In general, these additional scaffoldings are provided with the intent to help students figure out the 3 step nature of the quiz problem and the situation in which the CM/CME principle should be applicable, since these are the greatest difficulties students had as described in the previous section. Overall, the interviews suggest that simply directing students to go back to the solved problem and to explicitly identify the situation in which each physical principle is applicable in the solved problem didn't help students much partly because they didn't necessarily ask themselves why this principle was applicable here, and more importantly, why it wasn't applicable to some other situations. As noted earlier, even if students were able to describe that the CM principle was applicable in the solved problem from right before to right after the collision, they may still draw from their previous experience that the CM principle was always valid from $t=0$ to $t=\infty$. Thus, in the putty problem, they were likely to use the CM principle throughout.

The interviews suggest that a good way to help students learn the applicability of the CM/CME principles from the solved problem provided is to use the solved problem to guide students to contemplate deeply (1) why and how each principle is applied in the way shown in the solved situation, and (2) more importantly, why those principles are not applicable in other situations (e.g., why is it inappropriate to apply the CME or CM principle throughout the whole process). For example, the interviews with students E and F suggest that if a detailed discussion with the students about the solved problem was carried to help students learn deeply (e.g., about the fact that when the snowboard goes up the ramp, the net external force acting on the system is not zero and the velocity of the snowboard keeps changing, so the momentum couldn't be conserved), students are likely to realize themselves that in the quiz (transfer) problem, the momentum principle should be applicable only between right before and right after the collision. In comparison, in the interviews with other students, in which no similar discussion ensued, students were not be able to figure out how the CM principle should be involved to solve the putty problem unless an explicit instruction telling him/her to apply CM principle only right before and right after the collision was provided. A list of important issues for discussion related to the solution of the snowboard problem which may be beneficial for the students include: (a) When is the CM principle applicable? (The instructors may give students several choices including "from immediately before to immediately after the collision", "from the very beginning to the very end when the person reaches the maximum height", etc.) (b) Why isn't the CM principle applied elsewhere in the problem? Could it be applicable elsewhere in the problem? Why or why not? (c) From where to where in the problem is the CME principle applicable? Why do we apply CME only in this part of the problem? Can we go beyond that and apply it throughout the whole process in the problem? (d) What is an appropriate system for applying each of these principles? Such questions could keep students more actively involved in the learning process while working on the solved example and help them benefit more from the self-explanation process as discussed in Chi's study [68].

However, in order to solve the quiz (transfer) problem correctly, not only do students need to understand that neither CME or CM principle could be applied throughout the whole process, they also must have a clear picture of the 3 step structure of the quiz problem as well as how each sub-problem is connected to each other. In order to help students with these issues, in the interviews with students E and F, we also examined the effect of introducing a new problem (the "two block problem" shown in Appendix C) as a bridging problem to help students solve the putty problem involving three parts. This new two-block problem consists of only two steps: an object going down, colliding, sticking and moving together with another object on the horizontal part of the track. This bridging problem is a 2-step problem which is very similar to the solved problem except that the processes are reversed. After students E and F realized how to solve the new bridging problem correctly, we then asked them to take advantage of what they learned from these two problems to solve the 3-step putty problem. (A detailed description of the interventions students E and F received can be found in Table 5). We hypothesized that after the students understand how to solve the snowboard problem and the two block bridging problem, they will have a better idea of the three processes involved in the putty problem and they may be able to construct a holistic picture of how the different sub-problems should be connected.

Indeed, the interviews suggest that the additional scaffoldings provided to students E and F (which involves (a) a discussion guiding students to contemplate why and how each principle was applied in the way shown in the solution and not in other situations (b) the "two-block" problem) are very likely to benefit students. In the interviews with students A, C and D, in which the bridging problem was not provided and we didn't use the solved problem to discuss the issues related to the applicability of the CM/CME principle, we found that significant help from the researcher is usually required to help students solve the putty (transfer) problem correctly. For example, in addition to the reminder that the CE principle couldn't be applied in processes involving the inelastic collision, students C and D required an explicit instruction in "applying the CM principle only right before and right after the collision in the putty problem" as well as other assistances (such as guidance about how to connect the 3 subproblems to express $h_f$ in terms of $h_o$ or an instruction showing how to solve a particular subproblem involving CME) to help them solve the putty problem. Similarly, student A was not able to figure out the correct solution to the putty problem unless the whole problem was broken into sub-problems in which the target in each sub-problem was specified and other necessary scaffolding supports were provided.

On the other hand, in the interviews with students E and F, we found that after enough discussion with the students and explanation about why and how each principle was applied in the way shown in the solution to the snowboard problem, and after the students understood how to solve the bridging problem (i.e. the two-block problem) correctly following the discussion, they were able to take advantage of what they learned from the two problems (snowboard problem and the two-block problem) and correctly transfer their learning to solve the putty problem on their own. The critical scaffolding provided was the help in recognizing the similarities between the snowboard problem and the two-block problem and understanding how to solve the two-block problem correctly. One student required more help in understanding that something going down in the two-block problem is in principle the same as something going up in the solved snowboard problem. However, after students E and F recognized this similarity, understood that the mechanical energy of the system is not conserved during the inelastic collision, and realized why CM is not valid throughout the whole process in the snowboard problem or the two-block problem, the putty problem was not as difficult for them as for other students. After the scaffolding described above, without the help from the researcher, students E and F themselves recognized that the putty problem should be split into 3 sub-problems and CME, CM, and CME should be applied to the three consecutive sub-problems, respectively. Although the fact that the final velocity in one sub-problem becomes the initial velocity in the next sub-problem was somewhat frustrating for one student, both of them had a clear picture of the whole solution process, and were able to solve the quiz problem correctly on their own.

# IV. CONCLUSION AND FUTURE OUTLOOK

In this study, we found that it is very challenging for introductory physics students to transfer what they learned in the 2-step snowboard problem to solve the 3-step putty problem. In the calculus-based course, when students were asked to learn from the solution to the snowboard problem provided and take advantage of what they learned from the snowboard problem to solve the putty problem which is isomorphic, they didn't outperformed their peers in the comparison group, who were not provided with any scaffolding support to solve the putty problem. Although students in the algebra-based course who received the scaffolding with the solved problem significantly outperformed their peers in the comparison group, their average scores –which fell in the range of only 44% to 54% - suggest that there is still much room for improvement. Findings revealed that the greatest difficulty students had in transferring was in applying the physics principles learned from the solved problem to the new situation presented in the quiz problem in an appropriate way. Even though the solved problem could help students invoke the relevant principles in the quiz problem (which is the main reason why in the algebra-based course, students who received the scaffolding with the solved problem outperformed students in the comparison group, most of whom had no clue how to go about solving the putty problem), many students didn't have a clear plan for how those principles should be applied to solve the putty problem. They didn't understand how to decompose the putty problem into suitable sub-problems and they sometimes combined several processes into one, applied the physics principles in inappropriate situations, or applied the principles correctly but didn't discern their relevance to the final answer (target variable). For calculus-based students, many of them were able to invoke the relevant principles even without learning from the solved problem and the greatest difficulty was in applying the principles correctly. The scaffolding supports provided didn't help them significantly. Algebra-based students generally had even more difficulty in applying the principles correctly.

In a previous study, students were asked to perform analogical problem solving between a solved problem and an isomorphic quiz problem, both of which were two-step problems [46]. Comparing students' ability to transfer in this study to that in the previous study, we find that even though the problems in the previous study required the application of Newton's second law in the nonequilibrium situation, which is typically challenging for students, on average students displayed better transfer for the case discussed in the previous study. The fact that in our current study, the solved problem provided was a two-step problem whereas the targeted problem was a three-step problem made the transfer very challenging. With the existence of an additional step in the quiz problem, students could no longer map the solved problem directly to the quiz problem. They had to learn from the solved example and understand the circumstances for which each principle is applicable, so as to be able to systematically decompose the transfer problem into appropriate sub-problems (that can be dealt with one at a time with a single principle). The interviews suggest that students often superficially mapped the principles employed in the solved problem to the quiz problem without necessarily understanding the governing conditions underlying each principle or examining the applicability of the principles in the new situation in an in-depth manner.

Our findings suggest that determining the situations in which the CM and CME principles are applicable is very challenging for students despite providing them with a solved problem and more scaffolding is required in order to help students perform better on the transfer problem. However, the idea is not to spoon-feed them; rather, the dimensions of efficiency and innovation as described in Schwartz, Bransford and Sears's model are both important for transfer [26,65]. Students should be actively engaged in the analogical reasoning process themselves and in reconstructing, organizing and extending their knowledge structure. It is possible that if students are guided to think about the solution in more depth and contemplate the applicability of various principles in the solution carefully, they are more likely to benefit from the solved problem provided.

One possible way to guide students' self-explanation toward this goal is suggested in the interview part of this study and can be investigated in-depth in the future. In particular, if suitable questions are designed about the applicability of the principles used in the solved problem, and before students solve the quiz problem, they are asked to justify why those physics principles are applicable in the situations in the solved examples, but not in other situations, they may learn better from the solved problem provided.

Interviews also suggest that another strategy that may assist students in figuring out the 3 step structure of the putty problem is to add a bridging problem in this analogical problem solving activity. For example, after students learn from the solved snowboard problem, they can be asked to solve a bridging problem (such as the two-block problem discussed in the interview section) first before they solve the putty problem. In the interviews, we found that if the students understood why and how the CM and CME principles are applicable in the snowboard problem and the two-block problem, why the CM principle isn't valid for other sub-problems before or after the collision, and why the CME principle isn't applicable throughout the whole process in the two problems, they were likely to solve the putty problem correctly on their own without help. The bridging problem can be further tested in the future studies with a larger group of students to examine its effect in large classes.

Additional scaffoldings may also be designed and tested in the future studies to help students with specific difficulties in problem solving. For example, if the students are unsure about whether they should use the CM and CME principles for the inelastic collision process, instead of simply telling them the correct answer, an intervention could deliberately direct students to contemplate both the momentum and mechanical energy of the system right before and right after the collision (for example, by asking them to compute the speeds and kinetic energies at these two instances and compare whether their results are consistent with the predictions they made for both conservation laws). It is possible that by doing so, e.g., in the problem discussed in this study, they will be more likely to understand that some mechanical energy will be transformed into other forms of energy and they cannot simply set the initial potential energy of one putty equal to the final potential energy of both putties together without contemplating the inelastic collision process in between. Future studies could also explore transfer between a solved problem and an isomorphic transfer problem in situations where two students work together on making sense of the solved problem and on transferring their learning from one situation to another.

In summary, deliberately using an isomorphic worked out example to help students transfer what they learned from one context to another can be a useful tool to help students understand the applicability of physics principles in diverse situations and develop a coherent knowledge structure of physics. For introductory students, such well-thought out activities could provide a model for effective physics learning since the idea of looking at deep similarities beyond the surface features is enforced throughout these activities. Since our study suggests that it can be challenging for students to correctly apply what they learned from a 2-step problem to solve a 3-step problem, more scaffolding supports that are commensurate with students' prior knowledge may be required to help them realize the structure of the solution and to learn from the solved example effectively. It can be beneficial if the importance of looking for governing conditions underlying each principle and examining the applicability of the physics principles in the new situation in an in-depth manner are consistently explained, emphasized, demonstrated and rewarded by the instructors. It is possible that students will become more facile at the analogical problem solving processes and be able to transfer their learning from one situation to another if suitable scaffolding, practice and feedback are constantly provided to them throughout the whole course.

**Appendix A: The Solved Problem (Snowboard Problem)**

Your friend Dan, who is in a ski resort, competes with his twin brother Sam on who can glide higher with the snowboard. Sam, whose mass is 60 kg, puts his 15 kg snowboard on a level section of the track, 5 meters from a slope (inclined plane). Then, Sam takes a running start and jumps onto the stationary snowboard. Sam and the snowboard glide together till they come to rest at a height of 1.8 m above the starting level. What is the minimum speed at which Dan should run to glide higher than his brother to win the competition? (Dan has the same weight as Sam and his snowboard weighs the same as Sam's snowboard.)

**<Solution>**

**1. Description of the problem**

Knowns:

Dan's mass: $m_D = 60$ kg

The mass of Dan's snowboard: $m_B = 15$ kg

Desired minimum height : $h_{min} = 1.8$m

The distance between the initial position of the snowboard and the inclined plane: $D = 5$ m

Target quantity:

The minimum speed that Dan should run: $v_{D,min}$ ( If $v_D \geq v_{D,\,min} \Rightarrow h \geq h_{min}$ )

Diagram:

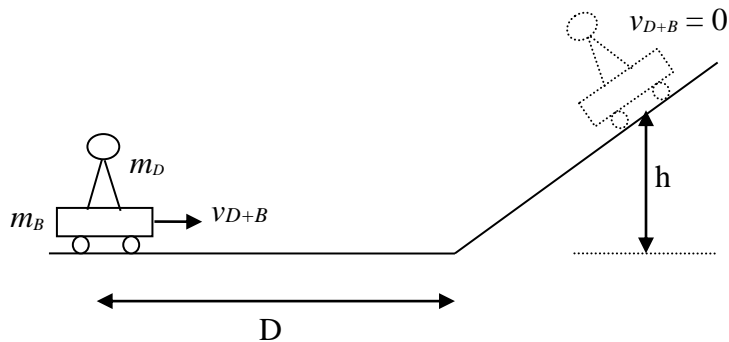

Assumptions:

Ignore retarding effects of friction and air resistance.

## 2. Constructing the solution

### Plan:

Suppose Dan runs with a speed $v_D$, and the height he reaches is $h$. If we can find $v_D$ in terms of $h$, then we can solve for $v_{D, min}$ for a desired $h_{min}$ given.

We notice that the problem has two distinct components:
* Dan jumping over the snowboard and coming to rest with respect to the snowboard in completely inelastic collision. We must find the speed of the snowboard with Dan in it after collision (assuming Dan's running speed is $v_D$ ).
* the system consisting of Dan and the snowboard go up the inclined plane and then stop at height h when the kinetic energy is zero.
* We note that we can use conservation of momentum for the first part to find the speed of the snowboard and Dan together.
* Then we can use conservation of mechanical energy for second part to find the height h at which the snowboard stops.

### Step B: Execution of the plan

**Sub-problem 1** – calculating speed $v_{D+B}$ of Dan and the snowboard after inelastic collision

Since the momentum of the system consisting of Dan and Snowboard is conserved

$$p_i = p_f$$

$$m_D v_D + m_B v_B = (m_D + m_B) v_{D+B}$$

Since the initial speed of the snowboard is zero,

$$m_D v_D = (m_D + m_B) v_{D+B}$$

$$v_{D+B} = \frac{m_D v_D}{(m_D + m_B)}$$

**Sub problem 2** – calculating the height reached before coming to a stop momentarily

From conservation of mechanical energy

$$\underbrace{KE_i + PE_i}_{\text{mechanical energy on horizontal surface}} = \underbrace{KE_f + PE_f}_{\text{mechanical energy on the incline when the snowboard stops momentarily}}$$

- $PE_i = 0$ if we choose the reference height to be on the horizontal surface
- $KE_f = 0$ at highest point since the speed there is zero

$$\therefore \frac{1}{2}mv_{D+B}^{2} = mgh \quad \text{where } m = m_D + m_B$$

$$v_{D+B}^{2} = 2gh$$

$$v_{D+B} = \sqrt{2gh}$$

$$\frac{m_D v_D}{(m_D + m_B)} = \sqrt{2gh}$$

$$v_D = \frac{m_D + m_B}{m_D}\sqrt{2gh}$$

$$v_{D,\,\text{min}} = \frac{m_D + m_B}{m_D}\sqrt{2gh_{\text{min}}}$$

$$v_{D,\,\text{min}} = \frac{60kg + 15kg}{60kg}\sqrt{2 \times 9.8\frac{m}{s^2} \times 1.8m}$$

$$= 7.4 \; m/s$$

**Final Result:** To win the competition, the minimun speed that Dan should run is 7.4 m/s.

### 3. Reasonability check of the solution:

- unit is correct for v
- In the limiting case $m_B = 0$, we expect $v_D = \sqrt{2gh}$, which agrees with our result

$$v_D = \frac{m_D + 0}{m_D}\sqrt{2gh} = \sqrt{2gh}$$

**Appendix B: The instruction provided to students in intervention group 3, and the quiz problem (putty problem)**

In this quiz, first browse over and learn from a solved problem, then solve the quiz problem below. Similar to the solved problem, this quiz problem below could be solved using conservation of energy and conservation of momentum. (You might have to use conservation of energy twice to find the height $h_f$ in terms of $h_0$.) Before you solve the quiz problem, go over the solved problem and then answer the following question.

**Quiz Problem:**

Two small spheres of putty, A and B, of equal mass, hang from the ceiling on massless strings of equal length. Sphere A is raised to a height $h_0$ as shown below and released. It collides with sphere B (which is initially at rest); they stick and swing together to a maximum height $h_f$. Find the height $h_f$ in terms of $h_0$.

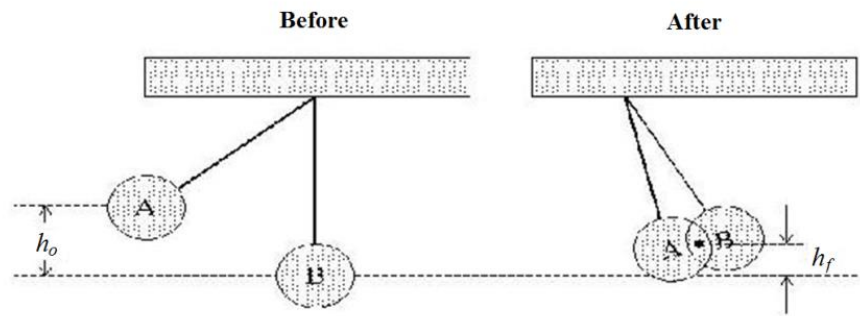

**Appendix C: The "Two-block" Problem Used in the Interview with Students E and F**

A block of mass $m_1$, initially at rest at a height of $h_o$, slides down a frictionless track (with an elevated end and a horizontal part as shown in the figure) and collides with another block of mass $m_2$. Then, the blocks stick and slide together on the horizontal surface. Find the speed of the blocks sliding together.

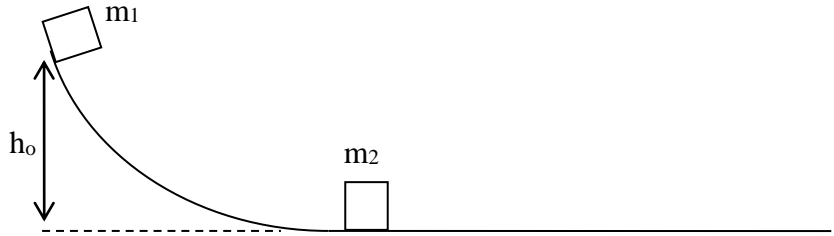

**Appendix D: Additional scaffoldings each interviewed student received (in addition to those used in the quiz in the introductory physics courses) and the effects of these scaffoldings**

<Student A>
In the interview, student A was first provided with the scaffolding of intervention 3 and was asked to follow the instruction in intervention 3 and to solve the putty problem while thinking aloud. The researcher didn't disturb him or provide any additional scaffoldings at this stage. His answer to the putty problem (after he had tried the problem to the best of his ability) is shown in Figure 6.

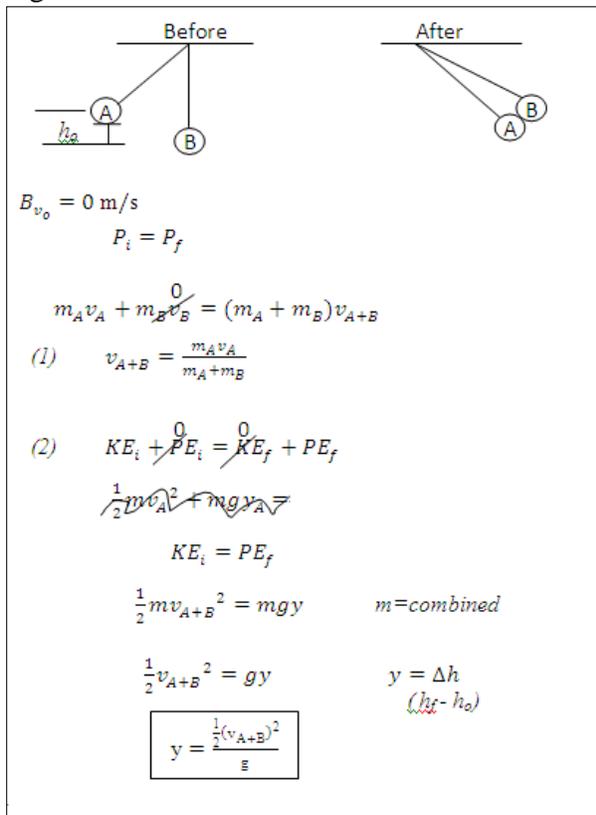

**Figure 6** Students A's answer to the putty problem when only the scaffolding support of intervention 3 was provided.

Since student A didn't successfully solve the putty problem (as shown in Figure 6), additional scaffolding supports were then provided to student A in order to help him solve the putty problem correctly. The additional scaffoldings that student A received and the effects of these scaffoldings are summarized in Table 13.

**Table 13. The additional scaffoldings that student A received and the effects of these scaffoldings.**

| Additional scaffolding provided to student A | Result of the scaffolding |
|---|---|
| Telling the student explicitly how to decompose the quiz problem into 3 parts (without telling him the goal of each sub-problem) | The student still struggled with the problem. |
| Telling the student that the goals of the 3 subproblems are to find the speed of sphere A right before collision, to find the speed of sphere A+B right after collision, and to find $h_f$, respectively. | The student was able to solve the 1st and 2nd subproblem correctly, but incorrectly wrote down $M_a g h_o = 1/2\ M_{ab} V_{ab}^2$ for the 3rd subproblem |
| Directing the student's attention to the fact that energy is not conserved in the inelastic collision. Only momentum is conserved. Then, providing in depth guidance to the student and demonstrating to him how to correctly solve the 3rd subproblem. | After the researcher demonstrated how to solve the 3rd subproblem, the student tried to reproduce it himself and felt that the problem started to make sense to him. |

<Student C>

In the interview, student C was first provided with the scaffolding of intervention 2 and was asked to follow the instruction in intervention 2 and to perform the task while thinking aloud. The researcher didn't disturb him or provide any additional scaffoldings at this stage. His answer to the putty problem (after he had gone through intervention 2 and tried the problem to the best of his ability) is shown in Figure 7.

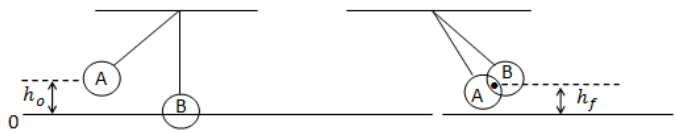

$$KE_i + PE_i = KE_C + PE_C$$

$$\frac{1}{2}m\cancel{v_i^2}^0 + mgh_o = KE_C + \cancel{PE_C}^0$$

$$mgh_o = \frac{1}{2}mv_C^2$$

$$\frac{2\cancel{m}gh_o}{\cancel{m}} = v_C^2$$

$$\sqrt{2gh_o} = v_C$$

$$KE_C + \cancel{PE_C}^0 = \cancel{KE_f}^0 + PE_f$$

$$\frac{\frac{1}{2}\cancel{m}v_C^2}{\cancel{m}g} = \frac{\cancel{m}gh_f}{\cancel{m}g}$$

$$\frac{1}{2}v_C^2 = h_f$$

$$h_o - h_f = h$$

$$h_o - \frac{1}{2}v_C^2 = h_{\text{in terms of } h_o}$$

**Figure 7** Students C's answer to the putty problem when only the scaffolding support of intervention 2 was provided.

Since student C didn't successfully solve the putty problem (as shown in Figure 7), additional scaffolding supports were then provided to student C in order to help him solve the putty problem correctly. The additional scaffoldings that student C received and the effects of these scaffoldings are summarized in Table 14.

**Table 14** The additional scaffoldings that student C received and the effects of these scaffoldings.

| Additional scaffoldings provided to student C | Results of the scaffolding |
|---|---|
| Directing the student's attention to the fact that energy is not conserved in the inelastic collision. Only momentum is conserved. | Student C decided to change his answer. He solved the 1st subproblem correctly and found that $v_i = \sqrt{2gh_o}$. But then he incorrectly applied the CM principle from "right before the collision" to the very end and was confused when he found that the two sides of the equation (one zero while the other is not) didn't equal each other |
| Directing the student's attention to the fact that the CM principle is only valid right before to right after the collision. | Student C was able to solve the 2nd and the 3rd subproblems correctly now, finding that $\frac{m_A v_i}{m_{A+B}} = v_f$ and that $\frac{\frac{1}{2}v_f^2}{g} = h_f$. However, he didn't know how to express $h_f$ in terms of $h_o$ (he didn't know what that means). He thought that $h_f$-$h_o$ may be the answer. |
| Explaining that "to express $h_f$ in terms of $h_o$" means "to express $h_f$ as a function of $h_o$" | Student C still struggled. |

| Guiding the student to connect the 3 subproblems in order to express $h_f$ in terms of $h_o$ (e.g., instructing him to plug one equation into another, guiding him to find out $\frac{m_A}{m_{A+B}}$ by considering the fact that $m_A=m_B$) | Student C got the correct answer. He also mentioned at the end of the interview that he struggled a lot because he had never encountered a problem which asked him to "express X in terms of Y". Instead, there were always number given that can be plugged into the equation. |
|---|---|

<Student D>

In the interview, student D was first provided with the scaffolding of intervention 2 and was asked to follow the instruction in intervention 2 and to perform the task while thinking aloud. The researcher didn't disturb him or provide any additional scaffoldings at this stage. His answer to the putty problem (after he had gone through intervention 2 and tried the problem to the best of his ability) is shown in Figure 8.

$$m_A v_A + m_B v_B = (m_A + m_B) v_{A+B}$$

$$\frac{1}{2}mv_o^2 + mgh_o = \frac{1}{2}mv_f^2 + mgh_f$$

$$\frac{\frac{1}{2}\cancel{m}v_o^2 + \cancel{m}gh_o - \frac{1}{2}\cancel{m}v_f^2}{\cancel{m}g} = h_f$$

$$h_f = \frac{\frac{1}{2}\cancel{v_o^2} + \cancel{g}h_o - \frac{1}{2}\cancel{v_f^2}}{\cancel{g}}$$

$$h_f = h_o$$

**Figure 8** Students D's answer to the putty problem when only the scaffolding support of intervention 2 was provided.

Since student D didn't successfully solve the putty problem (as shown in Figure 8), additional scaffolding supports were then provided to student D in order to help him solve the putty problem correctly. The additional scaffoldings that student D received and the effects of these scaffoldings are summarized in Table 15.

**Table 15 The additional scaffoldings that student D received and the effects of these scaffoldings.**

| Additional scaffoldings provided to student D | Effect of the scaffoldings |
|---|---|
| Directing the student's attention to the fact that energy is not conserved in the inelastic collision. Only momentum is conserved. | Student decided to change his answer. He felt that he should have solved the velocity of A+B like what was shown in the solved problem. He applied the CM principle and got the equation $v_{A+B} = \frac{m_A v_A}{(m_A+m_B)}$. However, he was confused because he thought that $v_A = 0$ |
| Directing the student's attention to the fact that the CM principle is only valid right before to right after the collision. | The student thought that the CM equation makes sense now. He then thought that he could plug in the velocity into a conservation of mechanical energy equation. Therefore, he wrote down $\frac{1}{2}mv_o^2 + mgh_o = \frac{1}{2}mv_f^2 + mgh_f, m = m_A$. He felt that the initial potential energy would be zero if the reference height was chosen at where sphere B was. But he was uncertain. He looked at the solution to the solved problem and found that in the solution, the initial velocity was the velocity after collision $v_{D+B}$, and that $v_f=0$ since it is at the highest position. However, he was still uncertain what he should do with his own CME equation. |
| Providing in depth guidance to the student in order to help him solve the subproblem (e.g, helping him to think through the equation by asking him to describe the situation in which the CME principle is applied and the meaning of each term in his CME equation; also, reminding him that $h_f$ is the maximum height) | Student D crossed out the $mgh_o$ term and the $\frac{1}{2}mv_f^2$ term, which led him to solve the 3rd subproblem correctly, for which he found that $\frac{\left(\frac{m_A v_A}{m_A+m_B}\right)^2}{2g} = h_f$<br><br>However, he hasn't successfully solved the putty problem because $h_f$ was not expressed in terms of $h_o$. |
| Reminding the student that he needs to find $h_f$ in terms of $h_o$ | Student D was not sure what to do and struggled a lot. |
| Telling the student that since his answer involves $v_A$, the speed of sphere A right before collision, he should try to relate $v_A$ to $h_o$. Also instructing him to focus only on putty A and ignore putty B while attempting to relate $v_A$ to $h_o$. | Student D felt that the CME principle should be used and wrote down $mgh_o = mgh_f$. Then he found that $v_A$ was not included in this equation and didn't know what to do. |
| Reminding him that the 1st step of a ball swinging down is similar to the 3rd step of two balls swinging together to the maximum height | Student D still struggled a lot |

| Reminding him that the CME principle should be applied only from the very beginning to right before the collision | After student D understood the applicability of the CME principle, he could solve the problem correctly. |

<Student E>

In the interview, student E was asked to first learn from the solved snowboard problem provided and then solved the "two-block problem" (with the solved problem in his hand) while thinking aloud. The researcher didn't disturb him or provide other support at this stage. After student E solved the "two-block problem" to the best of his ability, which is shown in Figure 9, more scaffolding supports were provided to student E in order to help him solve the "two-block problem" correctly. Moreover, a discussion of why the CM/CME principles can not be applied throughout the whole process in the solved problem was carried. After student E understood how to solve the "two-block" bridging problem correctly, he was then asked to take advantage of what he learned from the solved snowboard problem and the "two-block problem" to solve the putty problem. We found that after all these scaffolding supports, student E can solve the putty problem correctly without any help from the researcher. A more detailed description of the additional scaffolding support student E received (compared to those that the students received in the quiz) and the effects of these scaffoldings are summarized in Table 16.

$$m_1 gh + \frac{1}{2} m_1 v^2 = \cancel{m_{1+2} gh} + \frac{1}{2} m_{1+2} v^2$$

$$m_1 v_{10} + \cancel{m_2 v_{20}} = m_{1+2} v_f$$

$$m_1 v_{10} = m_{1+2} v_f$$

$$\frac{m_1 v_{10}}{m_{1+2}} = v_f$$

**Figure 9. Student E's solution to the "two-block problem" after learning from the solved snowboard problem provided.**

**Table 16. The additional scaffoldings that student E received and the effects of these scaffoldings.**

| Additional scaffoldings provided to student E | Effect of the scaffoldings |
| --- | --- |
| Asking the student to solve the "two-block" problem" first after learning from the solved snowboard problem provided. (Student E was not provided with the putty problem at this stage.) | Student E's answer to the "two-block" problem was shown in Figure 9. He noted that the CME equation was probably not needed since only the CM principle was used to came up with his final answer $\frac{m_1 v_0}{m_{12}} = v_f$. |
| Reminding the student that $v_o$ was not given in the problem statement | Student E decided to make use of the CME principle and came up with an answer shown in Figure 10. |
| Reminding the student that his CM equation didn't make sense if $v_o$ is the velocity of block A at height h, which should be zero. | Student E didn't know what to do. |
| Directing the student's attention to the fact that energy is not conserved in the inelastic collision. | Student E didn't know what to do. |
| Directing the student to go back to the solution to the solved problem and figure out the situation in which the CM/CME principles are applied in the solution. In particular, the researcher discussed with the student that the linear momentum cannot be conserved throughout the whole process in the solved problem since the initial velocity of the snowboard is in the horizontal direction, whereas the final velocity of the snowboard has a vertical component. | Student E was able to solve the "two-block problem" correctly and was able to successfully explain why neither the CM nor the CME principle can be applied throughout the "two-block" problem. Moreover, when student E was then asked to solve the putty problem, he solved it correctly without any difficulty. |

$$* \quad m_1 gh + \frac{1}{2}m_1 v_o{}^2 = m_{1+2}gh + \frac{1}{2}m_{1+2}v_f{}^2$$

$$m_1 v_{10} + m_2 v_{20} = m_{1+2} v_f$$

$$m_1 v_{10} = m_{1+2} v_f$$

$$\frac{m_1 v_{10}}{m_{1+2}} = v_f$$

$$\frac{1}{2}m_1 v_o{}^2 = \frac{1}{2}m_{1+2}v_f{}^2 - m_1 gh$$

$$v_o{}^2 = \frac{\frac{1}{2}m_{1+2}v_f{}^2 - m\,gh}{\frac{1}{2}m_1}$$

$$v_o = \sqrt{\frac{\frac{1}{2}m_{1+2}v_f{}^2 - m\,gh}{\frac{1}{2}m_1}}$$

$$\frac{m_1\left(\sqrt{\frac{\frac{1}{2}m_{1+2}v_f{}^2 - m_1 gh}{\frac{1}{2}m_1}}\right)}{(m_1 + m_2)} = v_f$$

**Figure 10. Student E's solution to the "two-block problem" after the researcher reminded him that $v_o$ was not given in the problem statement.**

<Student F>

In the interview, student F was asked to first learn from the solved snowboard problem provided and then solved the "two-block problem" and the putty problem (with the solved problem in his hand) while thinking aloud. The researcher didn't disturb him or provide other support at this stage. After student F solved the "two-block problem" and the putty problem to the best of his ability, which are shown in Figure 11 and Figure 12, respectively, more scaffolding supports were provided to student F in order to help him solve the "two-block problem" correctly. Moreover, a discussion of why the CM/CME principles can not be applied throughout the whole process in the solved problem was carried. After student F was able to solve the "two-block" bridging problem correctly, he was then asked to take advantage of what he learned from the solved snowboard problem and the "two-block problem" and tried to solve the putty problem a second time. We found that this time, student E can solve the putty problem correctly without any help from the researcher. A more detailed description of the additional scaffolding supports student E received (compared to those that the students received in the quiz) and the effects of these scaffoldings are summarized in Table 17.

$$mgh = \frac{1}{2}mv^2 \qquad\qquad m_2 \overset{0}{\|} PE$$

$m_1 \nearrow$

$$mgh = \frac{1}{2}(m_1 + m_2)v^2$$

$$v^2 = \frac{mgh}{\frac{1}{2}(m_1+m_2)}$$

$$v = \sqrt{\frac{mgh}{\frac{1}{2}(m_1+m_2)}}$$

**Figure 11. Student F's solution to the "two-block problem".**

| Before | | After |
|---|---|---|
| A | B | Both |
| PE= $mgh_o$ | 0 | $\frac{1}{2}(M)v^2 + mgh_f$ |
| KE= 0 | | |

Conservation of energy

$$M$$
$$\|$$
$$m_A gh_o = \frac{1}{2}(m_A + m_B)v^2 + (Mgh_f$$

$$h_f = \frac{m_A gh_o - \frac{1}{2}(M)v^2}{Mg}$$

**Figure 12. Student F's solution to the putty problem.**

**Table 17. The additional scaffoldings that student F received and the effects of these scaffoldings.**

| Additional scaffoldings provided to student F | Effect of the scaffoldings |
|---|---|
| Asking the student to solve the "two-block problem" as well as the putty problem after learning from the solved snowboard problem provided. | Student F's answers to the "two-block problem" and the putty problem are shown in Table 12 and Table 13, respectively. |
| Directing the student to describe what $v_f$ refers to in his solution to the putty problem and what value it should be | Student F decided that $v_f=0$ and his answer to the putty problem became $h_f = \frac{1}{2}h_o$ |
| Guiding the student to correctly solve the "two-block problem" by (1) directing his attention to the fact that energy is not conserved in an inelastic collision, (2) using the solved problem to help him understand that since the initial velocity of the snowboard is in the horizontal direction and the final velocity of the snowboard after it goes up the ramp has a vertical component, the linear momentum can not be applied throughout the whole process. It is conserved only right before and right after the collision | After Student F understood how to solve the "two-block problem", he could solve the putty problem correctly on his own. |